\documentclass[aps,twocolumn,floatfix,superscriptaddress]{revtex4-2}
\usepackage{graphicx}
\usepackage{dcolumn}
\usepackage{bm}
\usepackage{amssymb}
\usepackage{amsmath}
\usepackage{siunitx}
\usepackage{subfigure}
\usepackage{physics}
\usepackage{float}
\usepackage{multirow,tabularx}
\usepackage{sublabel}
\usepackage[utf8]{inputenc}
\usepackage{hyperref}
\usepackage{mathtools}
\usepackage{float}
\begin{document}


\title{Straintronics using the monolayer-Xene platform - a comparative study}
\author{Swastik Sahoo}
\affiliation{Department of Electrical Engineering, Indian Institute of Technology Bombay, Powai, Mumbai-400076, India}

\author{Namitha Anna Koshi}
\affiliation{Indo-Korea Science and Technology Center, Bengaluru-560064, India}


\author{Seung-Cheol Lee}
\affiliation{Electronic Materials Research Center, KIST, Seoul 136-791, South Korea}

\author{Satadeep Bhattacharjee}
\affiliation{Indo-Korea Science and Technology Center, Bengaluru-560064, India}

\author{Bhaskaran Muralidharan}
\email[E-mail:~]{bm@ee.iitb.ac.in}
\affiliation{Department of Electrical Engineering, Indian Institute of Technology Bombay, Powai, Mumbai-400076, India}

\begin{abstract}
Monolayer silicene is a front runner in the 2D-Xene family, which also comprises germanene, stanene, and phosphorene, to name a few, due to its compatibility with current silicon fabrication technology. 
Here, we investigate the utility of 2D-Xenes for straintronics using the \textit{ab-initio} density functional theory coupled with quantum transport based on the Landauer formalism. With a rigorous bandstructure analysis, we show the effect of strain on the K-point, and calculate the directional piezoresistances for the buckled Xenes as per their critical strain limit. Further, we compare the relevant gauge factors, and their sinusoidal dependences on the transport angle akin to silicene and graphene. The strain-insensitive transport angles corresponding to the zero gauge factors are $81^{\circ}$ and $34^{\circ}$ for armchair and zigzag strains, respectively, for silicene and germanene. For stanene as the strain limit is extended to $10\%$ and notable changes in the fundamental parameters, the critical angle for stanene along armchair and zigzag directions are $69^{\circ}$ and $34^{\circ}$ respectively. The small values of gauge factors are attributed to their stable Dirac cones and strain-independent valley degeneracies. We also explore conductance modulation, which is quantized in nature and exhibits a similar pattern with other transport parameters against a change in strain.  Based on the obtained results, we propose the buckled Xenes as an interconnect in flexible electronics and are promising candidates for various applications in straintronics.
\end{abstract}
\maketitle
\section{Introduction}
The prospect of exploiting the fascinating properties of graphene has initiated a paradigm shift towards the expeditious development of its analogous elemental monolayer sheets such as silicene~\cite{vogt2012silicene}, germanene~\cite{bianco2013stability}, stanene~\cite{osaka1994surface,zimmermann1997growth}, phosphorene~\cite{li2014black,liu2014phosphorene}, arsenene~\cite{kamal2015arsenene} to name a few. The robust electrical~\cite{novoselov2004electric,neto2009electronic}, mechanical~\cite{lee2008measurement,liu2007ab,kim2009large}, and chemical~\cite{geim2009graphene,rosas2011first} properties of graphene are touted to be the founding stone for investigation of Xenes as mentioned above and for the materials among the group $III$ and $V$ elements as well, which in due course will act as strong contenders for functional nano-devices~\cite{yuan2020recent}. Group $IV$ materials form a buckled structure ~\cite{molle2017buckled} whereas group $V$ materials can exist both in buckled and puckered structures ~\cite{yuan2020recent} as compared to the planar structure of graphene.\\
\indent Silicene, a monolayer of silicon (a group $IV$ element), is a prime contender for various applications due of its consonance with the already certified silicon industry ~\cite{tao2015silicene}. Adding to this advantage, silicene also possesses useful electro-mechanical properties ~\cite{houssa2010electronic,roman2014mechanical}. Germanene, the $2D$ counterpart of germanium, is also endowed with some extra-ordinary properties like silicene and they also possess a strong spin-orbit coupling~\cite{liu2011quantum,acun2015germanene}, giant magneto-resistance~\cite{rachel2014giant}, and a tunable bandgap due to applied electric fields~\cite{ni2012tunable}. Like graphene, silicene, and germanene exhibit linear energy dispersion relation near the Dirac points \cite{guzman2007electronic, cahangirov2009two}, zero band-gap~\cite{takeda1994theoretical}, and more importantly, they are dynamically stable~\cite{cahangirov2009two}. Among the $2D$ materials, stanene excels with outstanding properties and potential applications, though most of its electronic and structural properties are similar to other group $IV$ counterparts.
In group $V$ associates, black phosphorous is the most stable crystal structure among the allotropes which include red,  white, and violet phosphorous~\cite{hultgren1935atomic,thurn1966crystal}. This forms two-dimensional layers which are held together by weak van der Waals force~\cite{rodin2014strain} resulting in an orthorhombic puckered structure with a honeycomb like lattice arrangement in each layer ~\cite{brown1965refinement,island2015environmental}. This material can be singularized from other competitors due to its unique properties catering to various prospective listed as direct band gap from bulk to monolayer~\cite{tran2014layer} useful for optoelectronics~\cite{low2014tunable}, enhanced carrier transport mobility~\cite{li2014black,liu2014phosphorene} for nanoelectronics~\cite{xia2014rediscovering}, high on/off ratio for transistor applications ~\cite{li2014black,liu2014phosphorene}, in-plane anisotropy~\cite{low2014plasmons} related to electrical and optical properties, to name a few.   \\
\indent In spite of the above mentioned advantages, their respective applications in straintronics remain largely unexplored. In this paper, we delve into the strained lattice structure of 2D-Xenes monolayers in the quasi-ballistic regime, which corresponds to a length-scale of around $100$~nm - $200$~nm)~\cite{abidin2017effects}. The corresponding results are compared keeping in mind different nano electro-mechanical systems (NEMS)  applications and possible flexible electronics applications, which are quite diverse, satisfying the fundamental criterion of robust electronic and excellent mechanical response to strain~\cite{harris2016flexible}, followed by portability, and manufacturability ~\cite{wong2009materials}. \\
\indent The central goal of this paper is to compare silicene, germanene, stanene, and phosphorene, based on their fundamental properties, band structures, and transport properties in the presence of strain and explore different applications of Xenes like piezoresistivity, conductance modulation, etc., using \textit{ab-initio} density functional theory (DFT) and quantum transport theory. Specifically, we calculate the directional piezoresistance for different strain values along the armchair and zigzag directions. We typically obtaine a smaller value of the directional piezoresistances and their sinusoidal dependence. We show that the value of the gauge factor increases as we go up for group $IV$ elements along their atomic numbers in the periodic table. The strain-insensitive transport angles corresponding to the zero gauge factors are $81^{\circ}$ and $34^{\circ}$ for armchair and zigzag strains, respectively, for silicene and germanene. For stanene as the strain limit is extended to $10\%$ and notable changes in the fundamental parameters, the critical angle for stanene along armchair and zigzag directions are $69^{\circ}$ and $34^{\circ}$ respectively. The small gauge factor of buckled Xenes can be attributed to its robust Dirac cone and strain-independent valley degeneracy. Based on the obtained results, we propose the buckled Xenes as an interconnect in flexible electronics and are promising candidates for various applications in straintronics.\\
\indent The remainder of this paper is organized as follows. Section II represents a briefing about the theory and fundamental properties of the Xenes. Also, the tight binding parameters and band structures for the Xenes at different values of strains are displayed. Section III illustrates the methodologies to obtain the results. The depiction of these methods will go in parallel and corresponding applications have been discussed in detail. Results will be connoted forthwith, followed by a conclusion in section IV.
\section{Theory and properties} 
\begin{figure*}[t]
     \centering
     \includegraphics[width=1.08\textwidth]{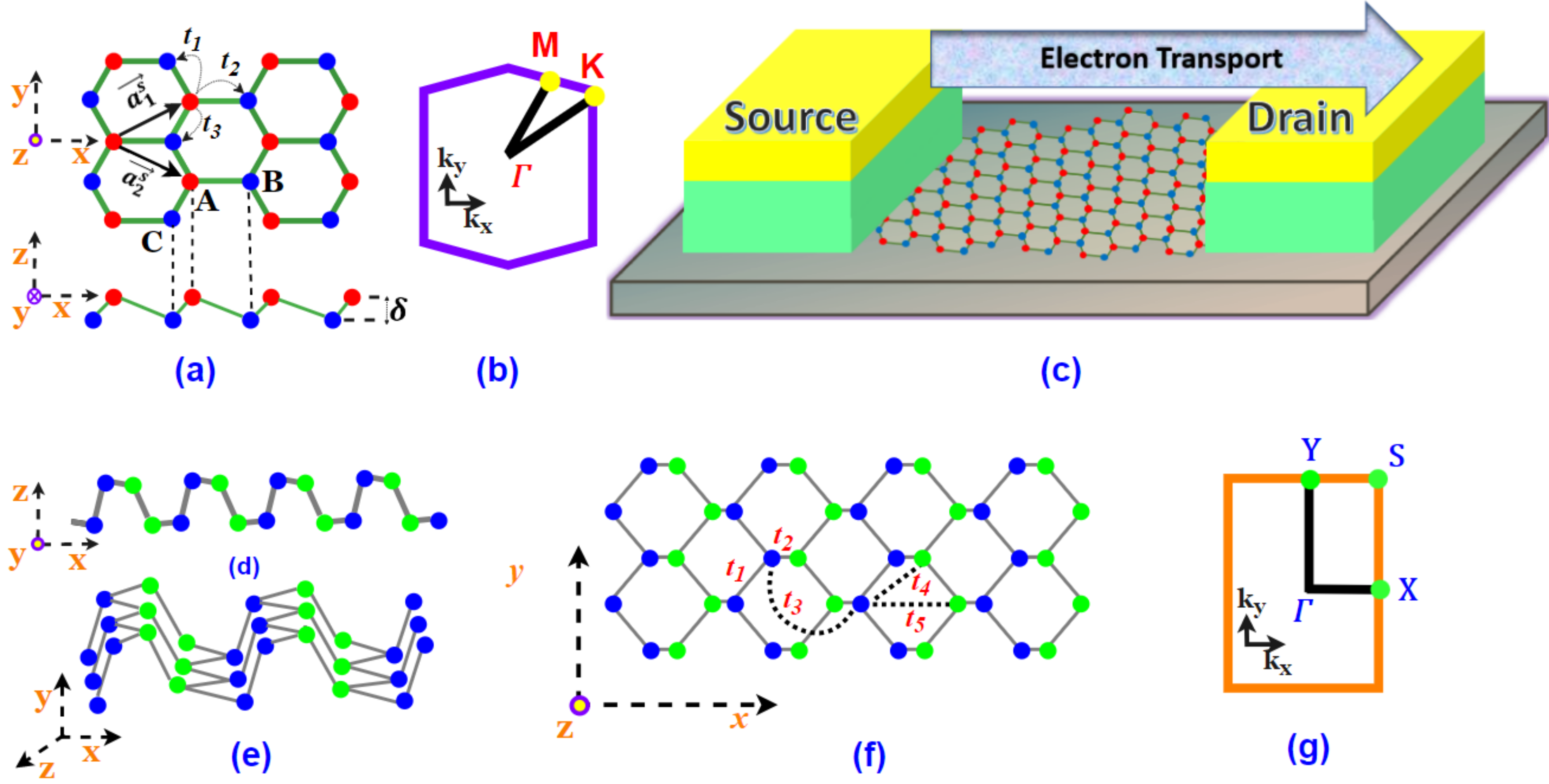}
     \caption{(a) Top view and side view of monolayer buckled Xene crystal lattice applicable to silicene, germanene, and stanene with $\protect\overrightarrow{a_1^{s}}$  and $\protect\overrightarrow{a_2^{s}}$ as the primitive vectors, and $\delta$ as the buckling constant. A and B represent the non-co-planar sub-lattices.  ${t_1}$, ${t_2}$ and ${t_3}$ represent the nearest neighbour tight binding parameters.(b) Schematic diagram depicting the high symmetry path ${\Gamma M K\Gamma}$ in the $1^{st}$ Brillouin zone of buckled strained Xene sheets. (c) Schematic presenting the setup and electron transport in Xene sheets. (d) Side view, (e) Perspective view, (f) Top view of phosphorene lattice. Blue- and -green-filled circles represent puckered-up and puckered-down phosphorous atoms, respectively. $t_1$ to $t_5$ represent the hopping parameters. (g) Sketch showing the high symmetry path ${Y \Gamma X S}$ in the $1^{st}$ Brillouin zone of phosphorene.}
    \label{P02_1}
\end{figure*}
In this section, we summarize the fundamental properties of Xenes in comparative form. However, we highlight the commonalities in the dependent and versatile properties of Xenes in the subsequent subsections.\par
\indent The lattice structure and their corresponding reciprocal symmetry, i.e., Brillouin zone, is shown in Fig.~\ref{P02_1}. Fig.~\ref{P02_1}a,~\ref{P02_1}b and ~\ref{P02_1}c are applicable for buckled Xenes corresponding to silicene, germanene, and stanene, respectively. The buckled honeycomb structure of Xenes consists of two triangular sub-lattices, shown in Fig. ~\ref{P02_1}a by red (denoted by A) and blue (denoted by B) dots, respectively. These sub-lattices are non-co-planar, which gives the Xenes a buckled honeycomb structure. The buckling constant $\delta$ differs for each Xene and is enlisted in table ~\ref{P02_table1} for reference. $t_1$, $t_2$, and $t_3$ represent nearest neighbor tight-binding (TB) parameters for buckled Xenes. Figure~\ref{P02_1}b depicts the first-Brillouin zone of Xenes in reciprocal space. For uniaxially strained Xenes, the high symmetry path is given by ${\Gamma M K\Gamma}$. Using $\textit{ab-initio}$ calculations, we obtain the band structures of strained Xene sheets along this path in the linear elastic regime~(see Fig.~\ref{P02_3}). The simulation setup for the calculation of gauge factor for all the monolayer buckled Xenes along different transport angles ($\theta$) is shown in  Fig.~\ref{P02_1}c. It consists of a buckled Xene sheet placed between two contacts, a source, and a drain. A uniaxial tensile strain (s) is applied along the armchair (AC) and zigzag (ZZ) direction varying from $0\%$ to $10\%$ depending on the material. The resistance is measured along each transport direction` $\theta$' ($0$$^{\circ}$ to $90$$^{\circ}$) for applied voltage ranging between $\pm 10$ mV. The electron transport direction is shown as a consequence of this process. The device dimension along the transport direction is nearly $100$~nm. Hence, the electrons undergo quasi-ballistic transport.\\
\indent The in-plane geometry of the black phosphorous crystal lattice is shown in Fig.~\ref{P02_1}d,~\ref{P02_1}e and ~\ref{P02_1}f. The side view and $3D$ view of the crystal lattice are shown in Fig.~\ref{P02_1}d and ~\ref{P02_1}e, respectively. Fig.~\ref{P02_1}f represents the alternate stacking of phosphorous atoms along the AC direction, giving rise to the puckered structure~\cite{rudenko2014quasiparticle}. The five hopping parameters, $t_1$, $t_2$, $t_3$, $t_4$, and $t_5$, are needed to describe phosphorene's low energy electronic structure within the TB model. $t_1$ and $t_2$ are the prominent nearest neighbor hopping parameters and $t_4$ accounts for the symmetricity between the conduction band and valence band. $t_3$ and $t_5$ are the minor parameters, used for dispersion correction~\cite{yang2016tunable}. The rectangle on $k_x$ and $k_y$ plane represents the two-dimensional Brillouin zone of the single puckered layer (see Fig.~\ref{P02_1}g). The high symmetry path is marked as ${Y \Gamma X S}$, and it is maintained throughout the calculation. \\
\indent A uniaxial tensile strain is applied along both AC and ZZ directions of buckled Xenes, which can be represented as~\cite{pereira2009tight}: 
\begin{equation}
s=s
  \begin{bmatrix}
    cos^2(\theta)-sin^2(\theta) & (1+\rho)cos(\theta)sin(\theta) \\
    (1+\rho)cos(\theta)sin(\theta) & sin^2(\theta)-cos^2(\theta)
  \end{bmatrix}
  \label{P02_eq1}
\end{equation}
Here, $\theta$ represents the transport angle, and it's default value is $0^{\circ}$ for ZZ strain and $90^{\circ}$ for AC strain. In this work, we have varied the transport angle from $0^{\circ}$ to $90^{\circ}$, and accordingly, the respective transport parameters are calculated.
\subsection{Fundamental properties}
We have listed the parameters related to the electronic and mechanical properties of all the Xenes in tables ~\ref{P02_table1} and ~\ref{P02_table2} subsequently. More parameters could have been documented, but we have kept only the relevant ones, keeping in mind the aspect of the paper. Graphene is not integral to this work but is listed as a reference material. Also, the dispersion relations for Xenes are expressed below.
\setlength{\tabcolsep}{10pt}
\begin{table*}[t]
\caption{Comparison of Lattice constant (a) in ($\AA$), buckling height ($\delta$), bond length (d) in ($\AA$), Spin-orbit coupling (SOC) in $meV$, Relaxation time near Dirac cone $(\tau_{d})$ in $ps$ , electron mobility ($\mu_e$) in $ 10^5 cm^2V^{-1}s^{-1}$, hole mobility ($\mu_h$) in $ 10^5 cm^2V^{-1}s^{-1}$ at $300$K for graphene, silicene, germanene, stanene, and phosphorene respectively. }
\centering
\begin{tabular}{c c c c c c c c}\\ 
 \hline\hline
 Material & a & $\delta$ & d & SOC & $\tau_{d}$ & $\mu_e$ & $\mu_h$ \\[0.08cm]
 \hline
 Graphene & 2.46~\cite{reich2002tight} & 0~\cite{neto2009electronic} & 1.42~\cite{neto2009electronic} & 0.05~\cite{yao2007spin} & 2.24~\cite{shao2013first} & 3.2~\cite{shao2013first} & 3.51~\cite{shao2013first}\\[0.03cm]
 Silicene & 3.83~\cite{qin2012first} & 0.44~\cite{lew2010hydrogen,cahangirov2009two} & 2.25~\cite{cahangirov2009two,qin2012first} & 1.55~\cite{liu2011quantum,yuhara2020beyond} & 1.84~\cite{shao2013first} & 2.57~\cite{shao2013first} & 2.22~\cite{shao2013first} \\[0.03cm]
 Germanene & 4.02~\cite{liu2011quantum} & 0.64~\cite{cahangirov2009two,qin2012first} & 2.38~\cite{cahangirov2009two,qin2012first} & 23.9~\cite{liu2011quantum,yuhara2020beyond} & 5.26~\cite{ye2014intrinsic} & 6.09~\cite{ye2014intrinsic} & 6.39~\cite{ye2014intrinsic} \\[0.03cm]
 Stanene  & 4.67~\cite{balendhran2015elemental} & 0.86~\cite{balendhran2015elemental} & 2.69~\cite{balendhran2015elemental} & 73~\cite{yuhara2020beyond}  & - & 27.7~\cite{nakamura2017intrinsic} & 40.1~\cite{nakamura2017intrinsic} \\[0.03cm]
 Phosphorene & 4.53~\cite{liu2014phosphorene,balendhran2015elemental} & 5.00~\cite{liu2014phosphorene,balendhran2015elemental} & 3.36~\cite{liu2014phosphorene} & 20~\cite{kurpas2018spin} & - & 0.011~\cite{qiao2014high} & 0.007~\cite{qiao2014high} \\[0.08cm]
 \hline\hline
\end{tabular}
\label{P02_table1}
\end{table*}

\setlength{\tabcolsep}{6.5pt}
\begin{table*}[htbp]
\caption{Comparison of elastic limit, isotropic limit, Poisson ratio $(\rho)$, Fermi velocity at Dirac point $(V_f)$ in $ (10^5 ms^{-1})$  at $300$K for graphene, silicene, germanene, stanene, and phosphorene respectively.}
\centering
\begin{tabular}{c c c c c}\\ 
 \hline\hline
 Material & Elastic limit & Isotropic region & $\rho$ & $V_f$  
 \\[0.08cm]
 \hline
 Graphene & $20\%$ (AC) and $26\%$ (ZZ)~\cite{liu2007ab,choi2010effects}  & $10\%$ (AC and ZZ)~\cite{pereira2009tight} & 0.14~\cite{neto2009electronic} & 6.3-14.4~\cite{xu2013graphene,nakamura2017intrinsic}  
 \\[0.03cm]
 Silicene & $15\%$ (AC) and $16\%$ (ZZ)~\cite{peng2013mechanical}  & $5\%$ (AC and ZZ) ~\cite{qin2014uniaxial} & 0.31~\cite{peng2013mechanical} & 5.1~\cite{xu2013graphene}  
 \\[0.03cm]
 Germanene & $16\%$ (AC) and $17\%$ (ZZ)~\cite{ding2018effects}  & $5\%$ (AC and ZZ~\cite{wang2013strain} & 0.33~\cite{john2016theoretical} & 3.8~\cite{xu2013graphene}  
 \\[0.03cm]
 Stanene  & $18\%$ (AC and ZZ)~\cite{mojumder2015mechanical} & $10\%$ (AC and ZZ)~\cite{van2014two} & 0.39~\cite{john2016theoretical} & 8.3~\cite{nakamura2017intrinsic}  
 \\[0.03cm]
 Phosphorene & $30\%$ (AC) and $27\%$ (ZZ)~\cite{peng2014strain}  & $10\%$ (AC and ZZ)~\cite{sa2014strain} & 0.20 (AC) and 0.70 (ZZ)~\cite{peng2014strain} & 3.5~\cite{fei2015topologically}  
 \\[0.08cm]
 \hline\hline
\end{tabular}
\label{P02_table2}
\end{table*}
\subsubsection{Buckled structured Xenes}
In their most stable form, silicene, germanene, and stanene form buckled hexagonal honeycomb structure~\cite{molle2017buckled} due to their larger bond length~\ref{P02_table1} compared to planar-like graphene having smaller bandgap. This large bond length makes the $\pi$ bond weaker leading to deviations from $sp^2$ hybridization. The energy dispersion relation applicable to all buckled Xenes is depicted as follows: 
\begin{equation} 
E(k)= \pm |t_2+t_1 e^{-(i \vec{k} \cdot \overrightarrow{a_1^{s}})}+t_3 e^{-(i\vec{k} \cdot \overrightarrow{a_2^{s}})}|.
\label{P02_eq2}
\end{equation}
Here, $\overrightarrow{a_1^{s}}$  and $\overrightarrow{a_2^{s}}$ represents the primitive lattice vectors, shown in Fig.~\ref{P02_1}a along the nearest neighbour tight binding parameters ($t_1$ to $t_3$).
\subsubsection{Puckered structured Xene}
  Phosphorene, on the other hand, forms a puckered structure ascribed to the covalent bond between each atom and three adjacent ones. This bond induces a strong anisotropy in phosphorene resulting in modulation of electronic, magnetic, optical, and transport properties~\cite{fei2014strain,wang2015electro}. Black phosphorous exists in four crystal structures: orthogonal, rhombic, simple cubic, and amorphous~\cite{wang2020present}. At room temperature, phosphorene has an orthorhombic crystal structure, as shown in Fig.~\ref {P02_1}.\\
 \indent The energy dispersion relation applicable to puckered Xenes ~\cite{yarmohammadi2020anisotropic} is represented as: 
\begin{equation} 
E(k)=f(k)\pm g(k).
\label{P02_eq3}
\end{equation}
where, $f(k)$ given by  $f(k)= 4t_4cos({k_{x}a}/{2})cos({k_{y}b}/{2})$ and the other component, $g(k)$ is depicted as: $g(k)= 2t_1 e^{-k_{x}a_{1x}}cos({k_{y}b}/{2})+ t_2e^{k_{x}a_{2x}} + 2 e^{k_{x}a_{3x}}cos({k_{y}b}/{2})+ t_5e^{k_{x}a_{5x}}$ . Here, $+(-)$ represents conduction (valence) band. \\ 
\indent The coefficients are given by $a_{1x}= 1.41763$ $\AA$ 
 and $a_{2x}= 0.79732$ $\AA$. These represent the intraplanar and interplanar distance for nearest-neighbour atoms along the x-direction, respectively. The other coefficients are expressed as: $a_{3x}=a_{1x}+2a_{2x}$, $a_{4x}=a_{1x}+a_{2x}$ and $a_{5x}=2a_{1x}+a_{2x}$. These parameters are obtained from the lattice structure of phosphorene ~\cite{yarmohammadi2020anisotropic} and $t_1$ to $t_5$ represent TB hopping parameters.
\subsection{Computational details}
The optimized band structure of silicene, germanene, stanene, and phosphorene are calculated from DFT using Vienna ab-initio Simulation Package (VASP)~\cite{kresse1996efficiency,kresse1996efficient,kresse1999ultrasoft}. The formalism of Perdew-Burke-Ernzerhof (PBE)~\cite{perdew1996generalized} within generalized gradient approximation (GGA) is used to treat exchange-correlation interactions. The buckled silicene, germanene, and stanene monolayers belong to the crystallographic space group $P-3m1 (164)$. A vacuum separation of $15 (\AA)$ is employed for germanene and stanene. Black phosphorene crystallizes in a puckered structure with space group $Pmna (53)$ different from the hexagonal Xenes. In the case of phosphorene, a vacuum of thickness $18 (\AA)$ is used in our calculations. Plane-wave cut-off energies within VASP are set to $500$ eV. Structures are optimized until each atom’s residual force becomes less than $0.01$ eV/(\AA). For geometry optimization of germanene and stanene, the Brillouin zone is sampled using a $41\times41\times1$ and $31\times41\times1$ k-mesh for 2-atom unit cell and 4-atom $1\times1$ rectangular super cell respectively. Similarly, for a monolayer of black phosphorus, a $31\times41\times1$ k-mesh is used for Brillouin zone sampling. The van der Waals correction is considered by considering Grimme’s semi-empirical DFT-D$2$ method~\cite{grimme2010consistent}. For germanene and stanene, a $1\times1$ rectangular supercell consisting of 4 atoms is constructed to study the effect of uniaxial strain on their electronic structures. Further calculation details are given in our previous work on silicene in \cite{sahoo2022silicene}. \\
\indent The strain range for different Xenes is chosen differently depending on their elastic limit and electronic properties, as listed in table~\ref{P02_table2}. The critical strain for each Xene can also be calculated quantitatively~\cite{zhang2011maximum}, based on the continuum mechanics model. The formula is given as follows:  
\begin{equation} 
S_{cr}= \frac{\pi^{2}t^{2}}{3(1-\rho^{2})W^2(\rho-(\frac{W}{2L})^2)}
\label{P02_eq4}
\end{equation}
    Here, $t$ is the thickness of the material, $L$ and $W$ represent the two-dimensional length and width of the material. $\rho$ represents Poisson ratio. In case of silicene, by taking its thickness to be 0.313 nm~\cite{peng2013mechanical} and the values of length, width, and Poisson ratio from ~\cite{sahoo2022silicene}, critical strain obtained is around $8\%$. In addition to this, there is a Dirac material-to-metal transition for a uniaxial tensile strain of $7.5\%$ for silicene ~\cite{peng2013mechanical}. So, the strain is restricted up to $7\%$ for silicene in our study. Similarly, for germanene, the study is restricted to a strain of $5\%$, attributed to the observed Dirac material-to-metal transition ~\cite{kaloni2013stability,ni2015electronic} and for stanene, upto $10\%$,  considering the range of isotropic elastic limit~\cite{shi2017ideal}.  
 \begin{figure*}[h]
     \centering
     \includegraphics[width=1.0\textwidth]{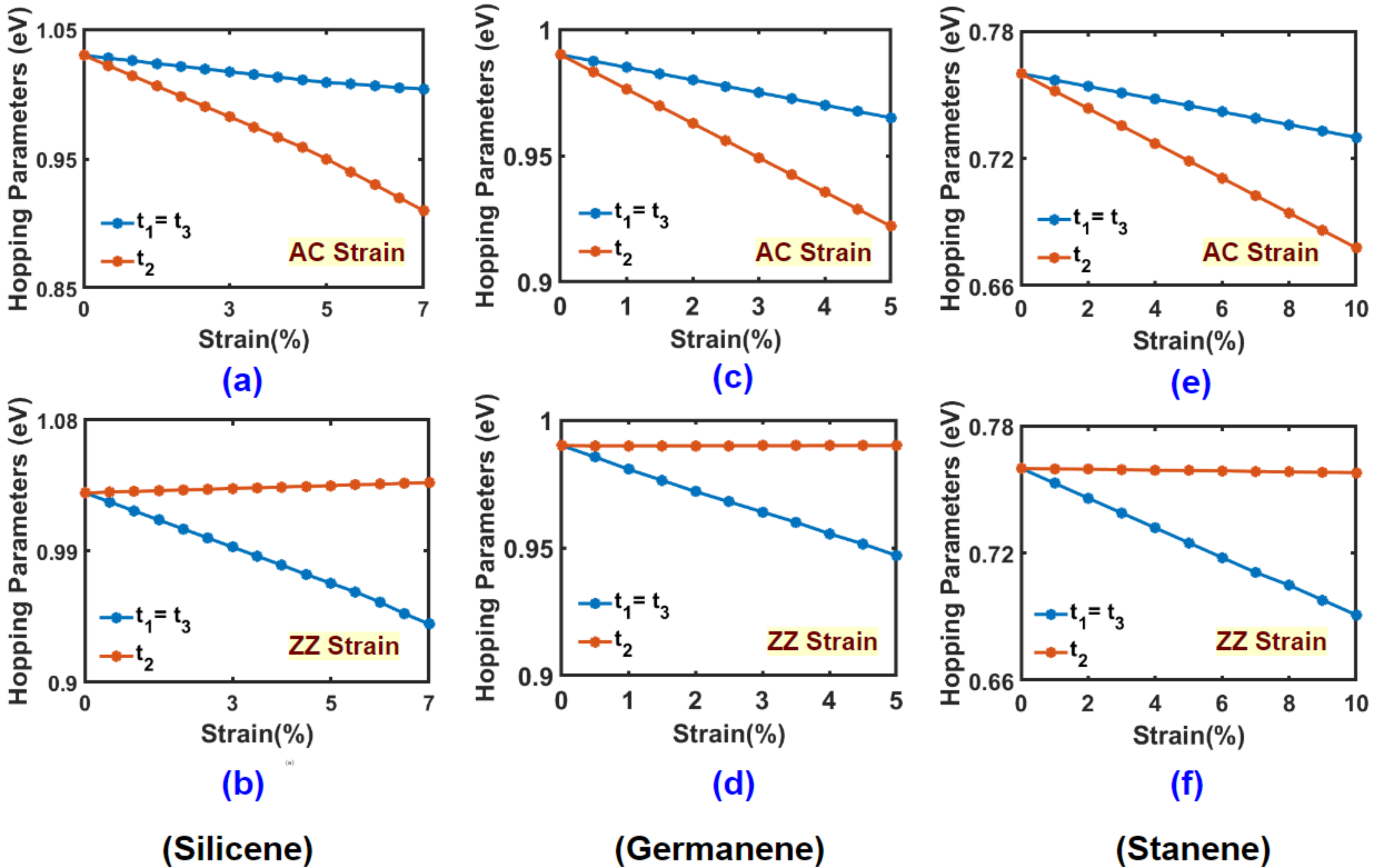}
     \caption{Variation of the nearest neighbor tight-binding parameters as a function of strain along the (a),(c),(e) armchair, and (b), (d), (f) zigzag directions for silicene, germanene, and stanene, respectively.}
    \label{P02_2}
 \end{figure*}
\begin{figure*}[t]
     \centering
     \includegraphics[width=0.98\textwidth]{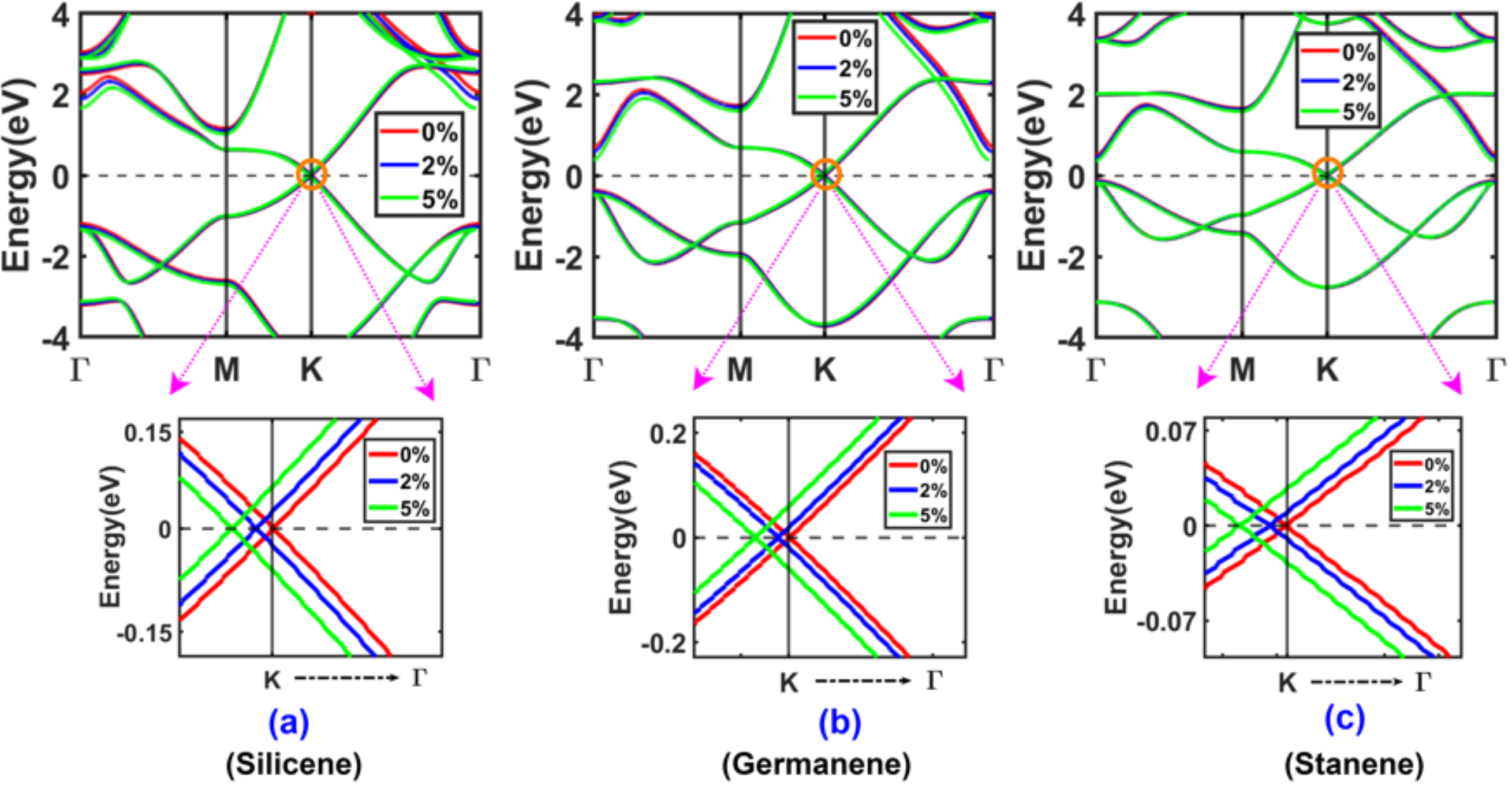}
     \caption{Band structure of a) silicene, b) germanene, c) stanene along the path $\Gamma MK \Gamma$ for AC strain adhering to the values of  $0\%$,  $2\%$ and  $5\%$ respectively. The magnified view of the point (orange encircled) shows the shifting of $K$ point away from the $K-\Gamma$ direction for each figure.} 
    \label{P02_3}
 \end{figure*} 
\subsection{Portrayal of hopping parameters and band structures }
This sub-section portrays the variation of nearest neighbor hopping parameters against the change in the values of strain for individual Xenes and the strain limit is dependent on their elastic limits and band structures.\\
 \begin{figure*}[t]
     \centering
     \includegraphics[width=1.0\textwidth]{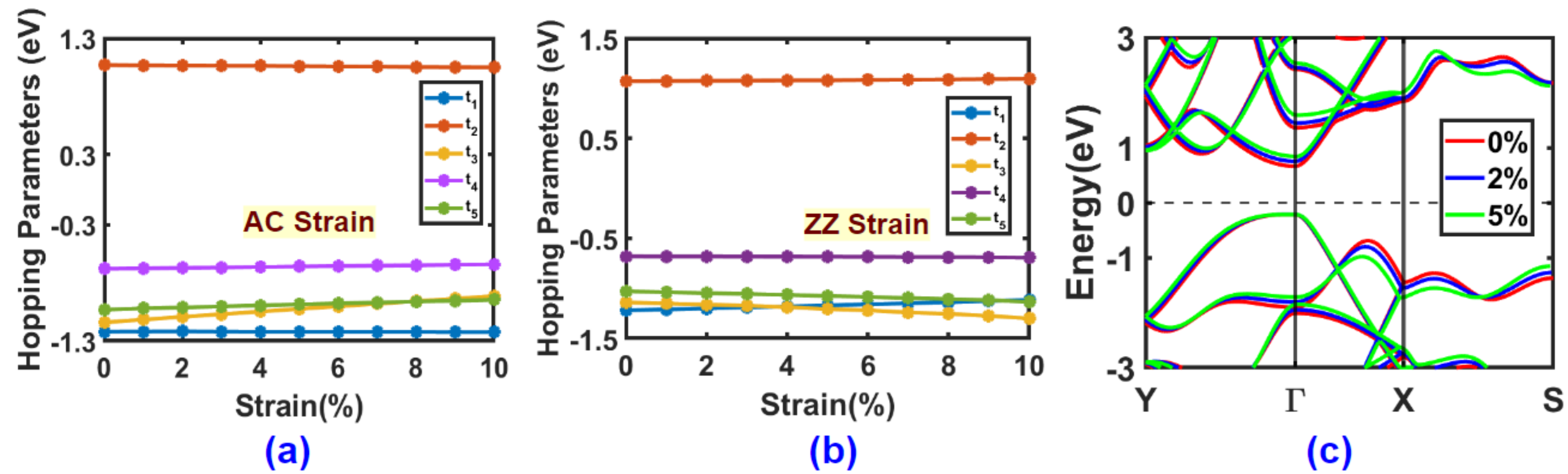}
     \caption{Variation of the nearest neighbor tight-binding parameters as a function of strain along (a)armchair and (b) zigzag directions for phosphorene, respectively. (C) The band structure of phosphorene along the path ${Y \Gamma X S}$ for AC strain adhering to the values of  $0\%$,  $2\%$, and  $5\%$.}
    \label{P02_4}
 \end{figure*}
 \indent We compute the Slater-Koster (SK) parameter, $V_{pp\pi}$, by fitting the DFT band structure to a tight binding (TB) model using TB-Studio \cite{nakhaee2020tight}. The relation between the SK parameter of strained and pristine Xenes is provided in~\cite{sahoo2022silicene}. The magnitude of hopping amplitude for pristine silicene, germanene, and stanene are $1.03$ $0.99$ and $0.76$ eV, respectively. These are obtained by benchmarking with our DFT results. In the case of black phosphorene, we use the 5-parameter tight-binding model of Rudenko and Katsnelson~\cite{rudenko2014quasiparticle} to describe the energy dispersion near the Fermi level. The hopping parameters $t_1$ to $t_5$ of pristine black phosphorene are obtained using Harrison’s method~\cite{harrison2004elementary}. The $\beta$ value is $2$ in the exponential relation linking the SK  parameter of strained and pristine black phosphorene~\cite{midtvedt2017multi}. The variation of hopping amplitudes as a function of uniaxial strain is presented in Fig.~\ref{P02_2} for silicene, germanene, and stanene ({$t_1, t_2, t_3$}) and Fig.~\ref{P02_4} for phosphorene ({$t_1, t_2, t_3, t_4, t_5$}). \\
 \indent From Fig.~\ref{P02_2}, it can be illustrated that all the monolayer Xenes follow the same pattern of hopping parameter variation along both AC and ZZ directions despite having different magnitudes. We obtain the tight-binding (TB) parameters for ten strain values irrespective of the elastic limit, and other values are taken from the line obtained by fitting. For phosphorene, we have shown the variation of hopping parameters with the applied uniaxial tensile strain, which is the first of its kind and one of the major highlights of the paper. From Fig.~\ref{P02_4}a and ~\ref{P02_4}b, it can be illustrated that the hopping parameters show minor variation compared to buckled Xene variation. Parameters $t_3$ and $t_4$ show slight variation with strain, whereas the rest of the parameters are almost constant with the strain variation irrespective of the direction.\\
 \indent The band structure of buckled and puckered Xenes computed from DFT calculations are shown in Fig.~\ref{P02_3} and Fig.~\ref{P02_4}c, respectively. The path shown in the band structures is in good agreement with the high symmetry path shown in the corresponding Brillouin zone. The region encircled in orange in each band structure shows the shifting of the K-point away from $\Gamma$-point with an increase in the uniaxial tensile strain value applied along the AC direction. This shifting of K-point is also satisfied quantitatively (see Supplementary material). The shifting of the K-point will be in the reverse direction when the strain is in ZZ direction~\cite{sahoo2022silicene}. The band structure of black phosphorene is presented in Fig.~\ref{P02_4}c. It is a direct band gap semiconductor with valence band maximum $(VBM)$ and conduction band minimum $(CBM)$ at $\Gamma$. The valence and conduction bands of black phosphorene in the $\Gamma-Y$ direction are flatter than the $\Gamma-X$ direction, which could lead to anisotropic effective mass in the AC and ZZ directions. With increasing tensile strain in the AC direction, the band gap increases for strain up to $7\%$.
Similarly, when subjected to increasing tensile strain in the ZZ direction, the band gap increases for strain up to $4\%$, after which it starts decreasing. This increase in bandgap is consistent with a previous report by Peng et al. \cite{peng2014strain}. Beyond $8\%$ tensile strain, the CBM shifts slightly away from the $\Gamma$-point. \\
\section{Results and discussion}
In this section, we obtain the piezoresistance gauge factor of silicene, germanene, and stanene along different transport angles using the theoretical models discussed alongside and rationalize our findings in terms of the change in transmission associated with the shifting and deformation of the Dirac cones. This section demonstrates the variation of quantized conductance as a function of strain. Further, we discuss the connotation of our results in flexible electronic devices.
\subsection{Piezoresistivity of Xenes}
The primary objective of this investigation is to calculate and compare the piezoresistivity of $2D$ Xenes in the nanoscale regime obtained from \textit{ab initio} calculation using DFT and $V-I$ characteristics obtained from the quantum transport model. The procedure to obtain the hopping parameters is discussed in the previous section, and the transport model will be discussed here. Phosphorene has a bandgap of $1.52$ eV~\cite{yarmohammadi2020anisotropic} at zero strain. So, the band counting method is not applicable here as it does not form a zero band gap or linear energy dispersion relation at $CBM$ and $VBM$. Another process to calculate the piezoresistance of these kinds of materials is the NEGF method, which has already been calculated by Naurbaksh \textit{et'al}~\cite{nourbakhsh2018phosphorene}. However, we will compare the data obtained using this method. Henceforth, our calculation has been restricted to germanene and stanene. The results of silicene~\cite{sahoo2022silicene}, germanene, and stanene will be comprehensively compared in the subsequent subsection.
\begin{figure*}[h]
     \centering
     \includegraphics[width=0.92\textwidth]{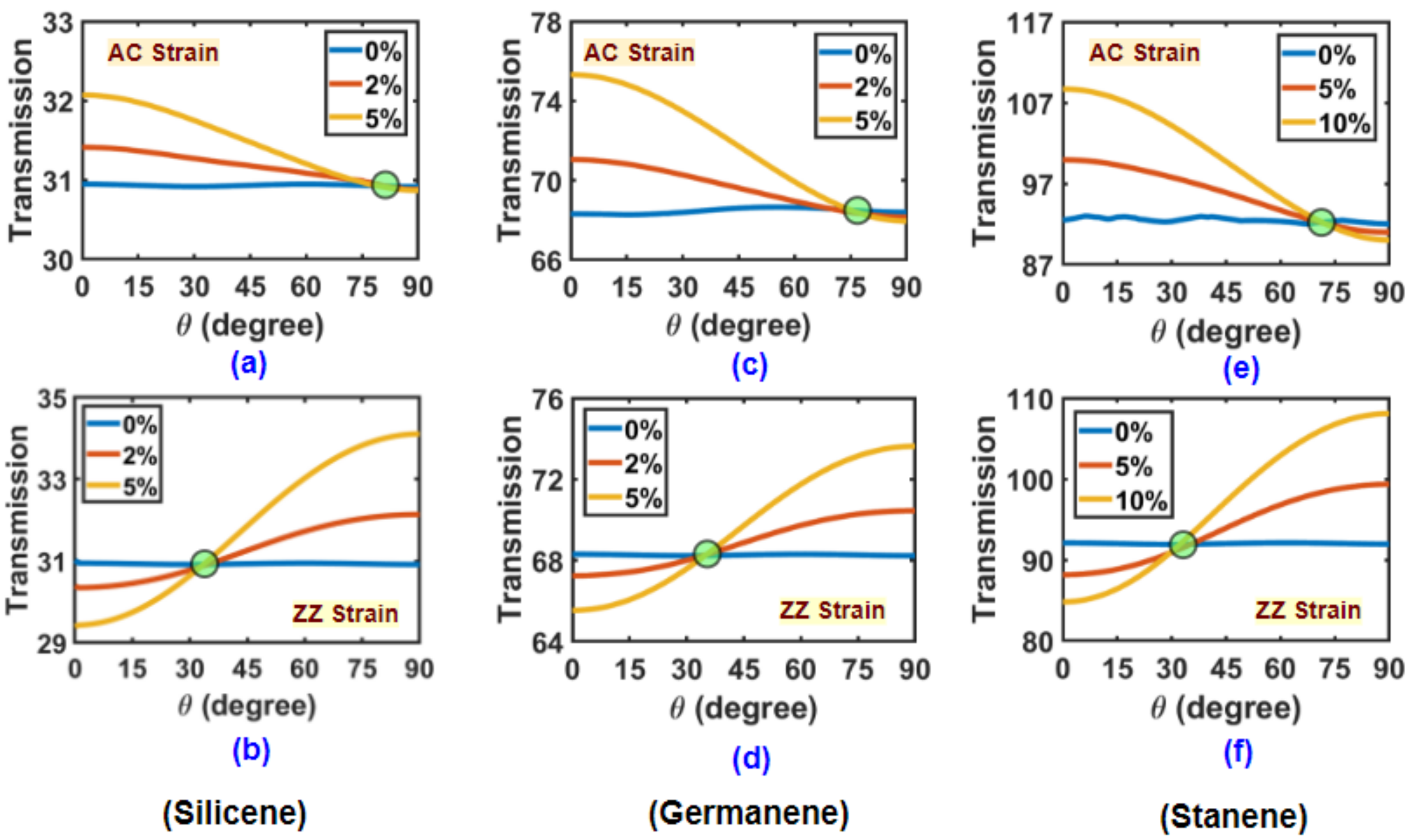}
     \caption{The plots of transmission as a function of transport angle $(\theta)$ for strain values $0\%$,  $2\%$ and  $5\%$ along the (a),(c) armchair, and  (b), (d) zigzag directions for silicene and germanene respectively. Transmission plots for stanene akin to the strain values of $0\%$,  $5\%$, and  $10\%$ along (e) armchair and (f) zigzag directions. The green circle in all the plots represent the critical angle.}
    \label{P02_5}
 \end{figure*}
 
  \begin{figure*}[h]
     \centering
     \includegraphics[width=0.92\textwidth]{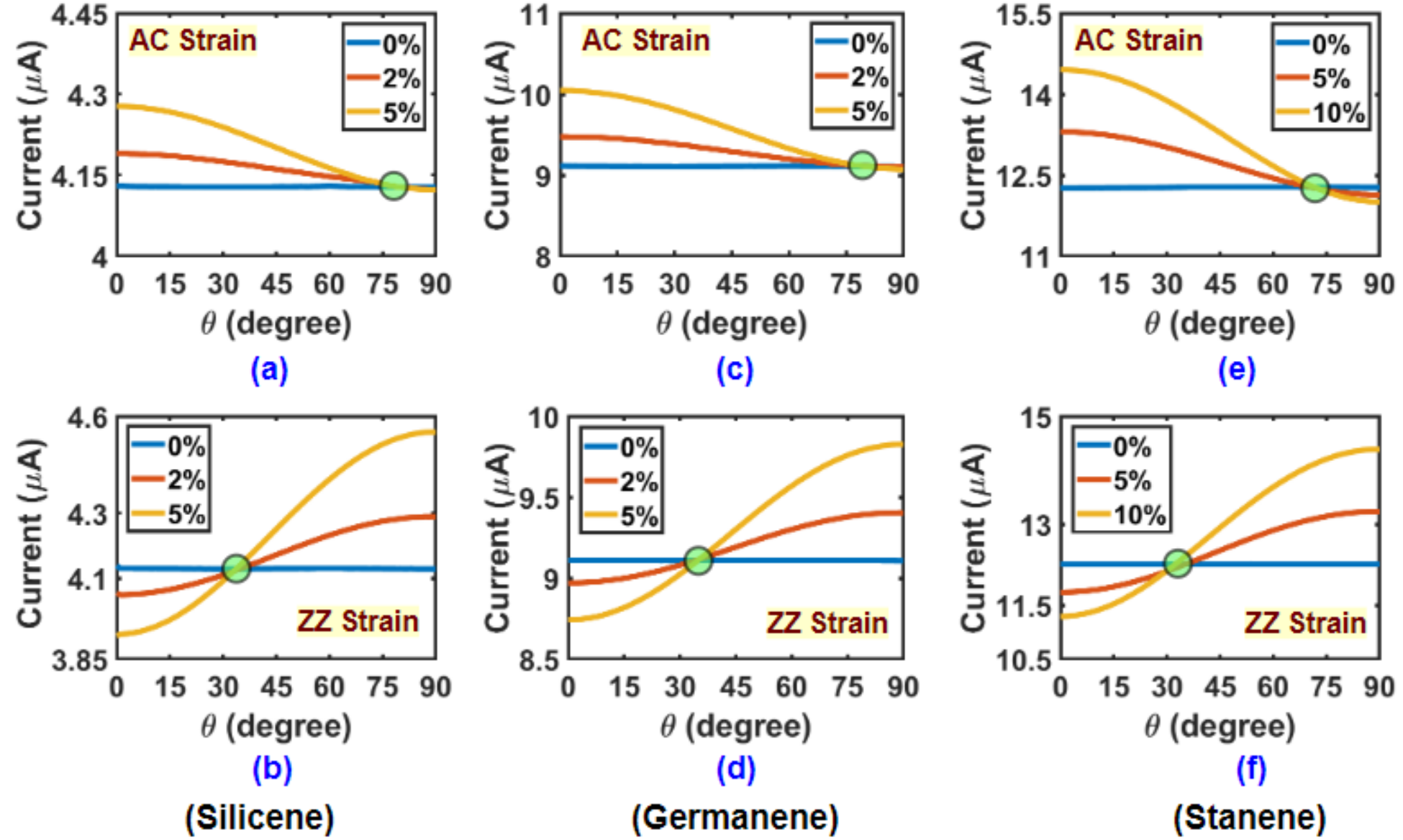}
     \caption{The plots of current as a function of transport angle $(\theta)$ for strain values $0\%$,  $2\%$ and  $5\%$ along the (a),(c) armchair, and  (b), (d) zigzag directions for silicene and germanene respectively. Current plots for stanene akin to the strain values of $0\%$,  $5\%$, and  $10\%$ along (e) armchair and (f) zigzag directions. The green circle in all the plots represent the critical angle.}
    \label{P02_6}
 \end{figure*}
\indent After procurement of TB parameters, we use the Landauer formula~\cite{datta2018lessons} to calculate the current-voltage characteristics of silicene in the quasi-ballistic regime. This method is common to all the Xenes discussed here. The Landauer formula is expressed as:
\begin{equation} 
 I_\theta^{s} (V)=\frac{2q}{h} \int_{-\infty}^{\infty} {T^s}(E) [f_1(E-\mu_1)-f_2(E-\mu_2)] dE
\label{P02_eq5}
\end{equation}
where $f_{1}$ and $f_{2}$ are the Fermi functions, and $\mu_{1}$ and $\mu_{2}$ are the electro-chemical potentials at the source and drain ends respectively, $T^{s}$ is the transmission with strain in the Xenes, $q$ is the electronic charge, $h$ is the Planck's constant, $E$ is the energy well within the linear regime, and  $I_\theta^{s} (V)$ is the current along the transport direction $(\theta)$ at an applied strain. Here, the voltage varies from $-10$~mV to $+10$~mV. In general, the strained transmission $T^{s}(E)$ is the product of transmission probability $T(E)$ and mode density $M_\theta^{s}(E)$. The transmission probability depends on both the length $(L_s)$ of the Xene sheet and their mean free path $(\lambda)$ in a quasi-ballistic regime. The mean free path in the linear regime depends on the relaxation time near the Dirac cone~\cite{shao2013first}, and as the calculation is in a quasi-ballistic regime, this will be nearly equal to the length of the Xene sheet~\cite{abidin2017effects}. The fundamental parameters in the ballistic regime are independent of this mean-free path. Both regimes' parameters are compared and tabulated in table~\ref{P02_table5}. The mean free path for silicene and germanene are calculated from ~\cite{abidin2017effects,gaddemane2016theoretical}, and the same method is applied for stanene to get the value. These values are satisfied in the context of a quasi-ballistic regime. $L_s$ represents the length of the strained Xene sheets, and with strain, its relation is depicted as $(L_s)=L(1+s\rho)$. The transmission probability is given as follows:
\begin{equation} 
T(E)=\frac {\lambda (E)}{\lambda(E)+L_s},
\label{P02_eq6}
\end{equation}

 \begin{figure*}[t]
     \centering
     \includegraphics[width=0.95\textwidth]{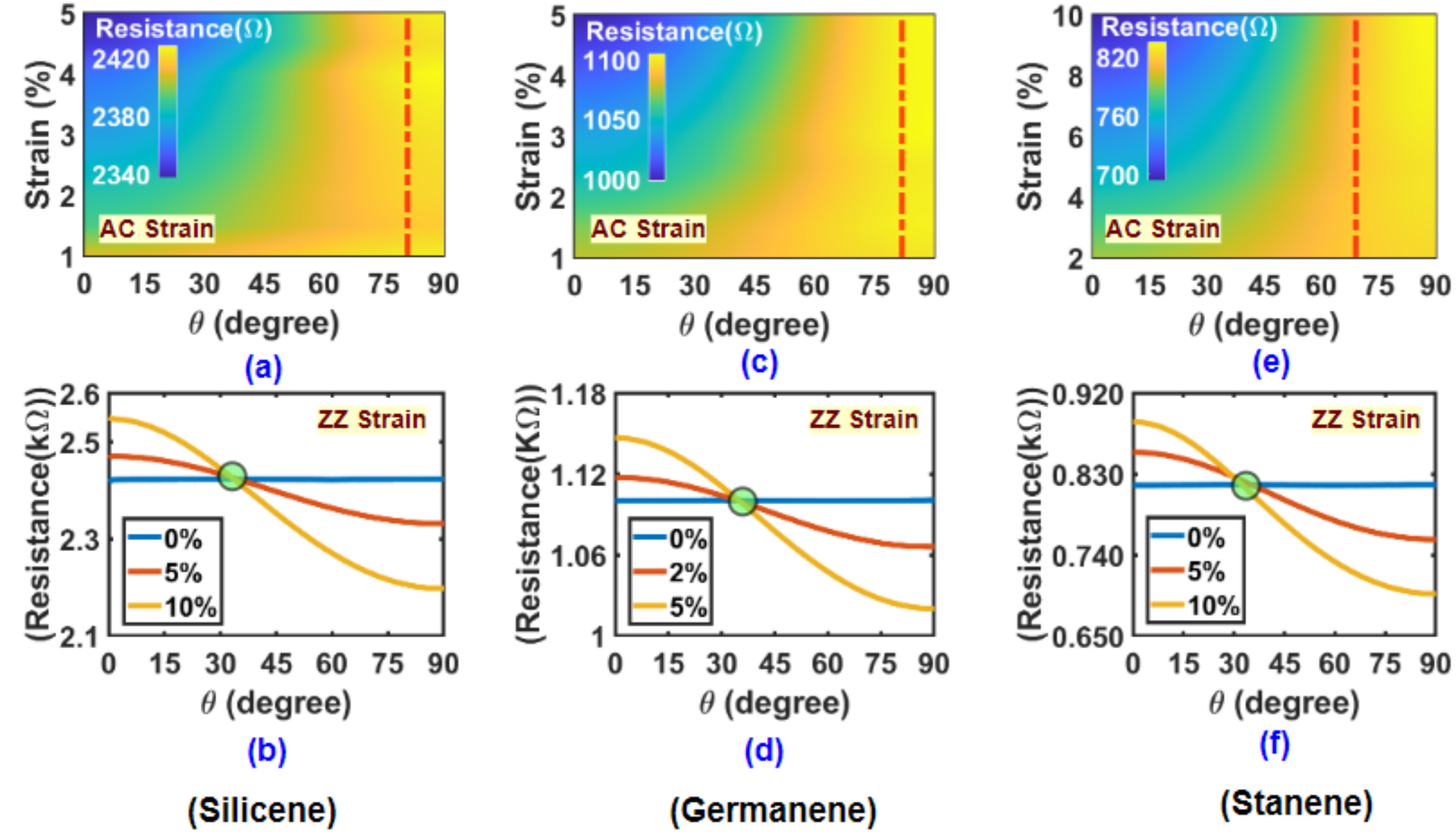}
     \caption{The plots of resistance as a function of transport angle $(\theta)$ for strain values $0\%$,  $2\%$ and  $5\%$ along the (a),(c) armchair, and  (b), (d) zigzag directions for silicene and germanene respectively. Resistance plots for stanene akin to the strain values of $0\%$,  $5\%$, and  $10\%$ along (e) armchair and (f) zigzag directions. The red dotted line in the top row and green circle in the bottom row for each figure represents strain insensitive transport angle.}
    \label{P02_7}
 \end{figure*}
 The transmission for silicene and germanene as a function of $(\theta)$ for $0\%$, $2\%$ and $5\%$ AC strain is shown in Fig.~\ref{P02_5}a and ~\ref{P02_5}c and for ZZ strain in Fig.~\ref{P02_5}b and ~\ref{P02_5}d respectively. The transmission of stanene is shown for the strain values of $0\%$, $5\%$ and $10\%$ in Fig.~\ref{P02_5}e and ~\ref{P02_5}f for AC and ZZ strain, respectively. Here, $\theta$ varies from $0$$^{\circ}$ to $90$$^{\circ}$. From the figures, we infer that all materials will exhibit opposite patterns concerning AC and ZZ strain. Mode density at the Dirac cone is required to calculate the effective number of transverse modes passing through the Dirac point, which is accomplished through the band-counting method ~\cite{sinha2019piezoresistance,sinha2020graphene}. The Dirac cone degeneracy for silicene in the first Brillouin zone is two~\cite{liu2013dirac,sinha2019piezoresistance}. So, effectively the mode density is twice the number of transverse modes passing through the Dirac cone. The pictorial representation of this calculation is given in the supplementary material. Mathematically, mode density can be expressed as $ M_\theta^s(E)= 2\times n_\theta^s(E)$. Here, $n_\theta^s(E)$ is the number of transverse modes passing through the constant energy surfaces of the Dirac cone at energy `$E$'. The superscript '$s$' indicates transverse modes are counted with applied strain on the Dirac cone. These transverse modes have a separation of $2\pi/w_s$, where $w_s$ represents the width of strained Xene sheet and is equal to  $w_s= w(1+\frac{s}{100})$. Here $w$ represents the width of the unstrained Xene sheet. Thus, transmission with strain is computed as: $T^{s}(E)= T(E)*M_\theta^s(E)$ as explained before.\\
 \indent The transmission remains constant at zero strain for varying transport angles due to circular constant energy surfaces in the Dirac cone of materials. Due to AC strain, the value of $L_{x}$ (length of Dirac cone measured along the x-axis) increases while $L_{y}$ (length of Dirac cone measured along the y-axis) decreases. As a result, a circle like the Dirac cone is deformed into an ellipsoid structure, and the number of transverse modes passing through this will increase because of the large surface area, which increases the mode density and eventually increases transmission. So, the transmission has a higher value at $\theta=$ 0$^{\circ}$ than unstrained Xenes sheets, and it decreases with an increase in the transport angle due to the change in the angle of the transverse modes propagating through the Dirac cone. On the contrary, the value of $L_{x}$ decreases while $L_{y}$ increases for ZZ strain which deforms the Dirac cones into an oval shape, and mode density will reduce here.
For the above same reason, the transmission along $\theta=$ $0^{\circ}$ has a lower value than the transmission of unstrained Xenes. However, due to a change in the angle of the propagating modes, the transmission increases with an increase in $\theta$. \\
 \indent The plots of current for silicene and germanene as a function of transport angle for $0\%$, $2\%$ and $5\%$ AC strain is shown in Fig.~\ref{P02_6}a and ~\ref{P02_6}c and for ZZ strain in Fig.~\ref{P02_6}b and ~\ref{P02_6}d respectively. The current plot for stanene is shown for the strain values of $0\%$, $5\%$ and $10\%$ in Fig.~\ref{P02_6}e and ~\ref{P02_6}f for AC and ZZ strain, respectively. The plots will depict similar behavior to the transmission plots as current is directly proportional to the transmission at a particular energy [see Eq.~\eqref{P02_eq5}]. The green bubble in both the transmission and current plot indicates an intersection point, where the strained and unstrained lines of transmission and current coincide at a single value in their respective plots, and the corresponding intersection angle is called the critical angle. The critical angle is the angle at which the material is insensitive to the applied strain. The transmission and value of current increases as we go down group $IV$ in the periodic table, which can be ascribed to the increase in radius of the Dirac cone (see Table.~\ref{P02_table5}). The above line can be a general statement applicable to group $IV$ elements.\\
\indent The contacts are assumed to be ideal in our calculations. The energy range for this Fermi-Dirac distribution is from $-0.2$ eV to $+0.2$ eV, which is well within the linear regime. Once current in eq.~\eqref{P02_eq5} is obtained, the piezoresistance is easily calculated from the dynamic change in voltage and current values. Therefore, it is depicted as follows: 
\begin{equation} 
R_\theta^{s}=\frac{dV}{dI_\theta^{s} (V)}
\label{P02_eq7}
\end{equation}
The above equation is used to calculate the piezoresistance at different transport angles. The plots of resistance versus $\theta$ as a function of AC strain are shown in Figs.~\ref{P02_7}a,~\ref{P02_7}c and ~\ref{P02_7}e for silicene, germanene, and stanene respectively. Similarly, for the ZZ strain, the plots are shown in Fig.~\ref{P02_7}b, ~\ref{P02_7}d and ~\ref{P02_7}f, respectively. It is obvious that resistance will show an opposite pattern concerning current plots. However, the critical angle remains constant for resistance as well, and the same can be depicted in the corresponding diagram of AC strain through a significant red line drawn vertically. The value of resistivity obtained in this work for $1~\mu m$ wide Xene sheets in the quasi-ballistic regime are \SI{2.42}{\kilo\ohm}, \SI{1.092}{\kilo\ohm}, \SI{0.815}{\kilo\ohm} for silicene, germanene and stanene respectively (refer Table~\ref{P02_table5}). It is now quite obvious that as the radius of the Dirac cone increases, the corresponding resistance value will go down. Using these values, we can calculate Angular Gauge Factor (AGF). AGF is expressed as: 
\begin{equation} 
(AGF)_\theta^{s}=\frac{(R_\theta^{s}-R_\theta^{0})}{s*R_\theta^{0}},
\label{P02_eq8}
\end{equation}
where, ${R_\theta}^{0}$ is the resistance at zero strain and ${R_\theta}^{s}$ is the resistance at a particular value of strain $s$.
As the strain is applied throughout the transport angle $(\theta)$, the AGF is the average of all GFs at different strains along a particular transport angle. Thus, the average AGF is expressed as:  
\begin{equation} 
(AGF)_\theta= \overline {(AGF)_\theta^{s}}
\label{P02_eq9}
\end{equation}
The GFs of all the monolayer Xenes in the AC and ZZ strain are shown in Fig.~\ref{P02_9}. The result will be discussed in the subsequent subsection.
\subsection{Conductance Modulation}
\begin{figure*}[t]
     \centering
     \includegraphics[width=0.95\textwidth]{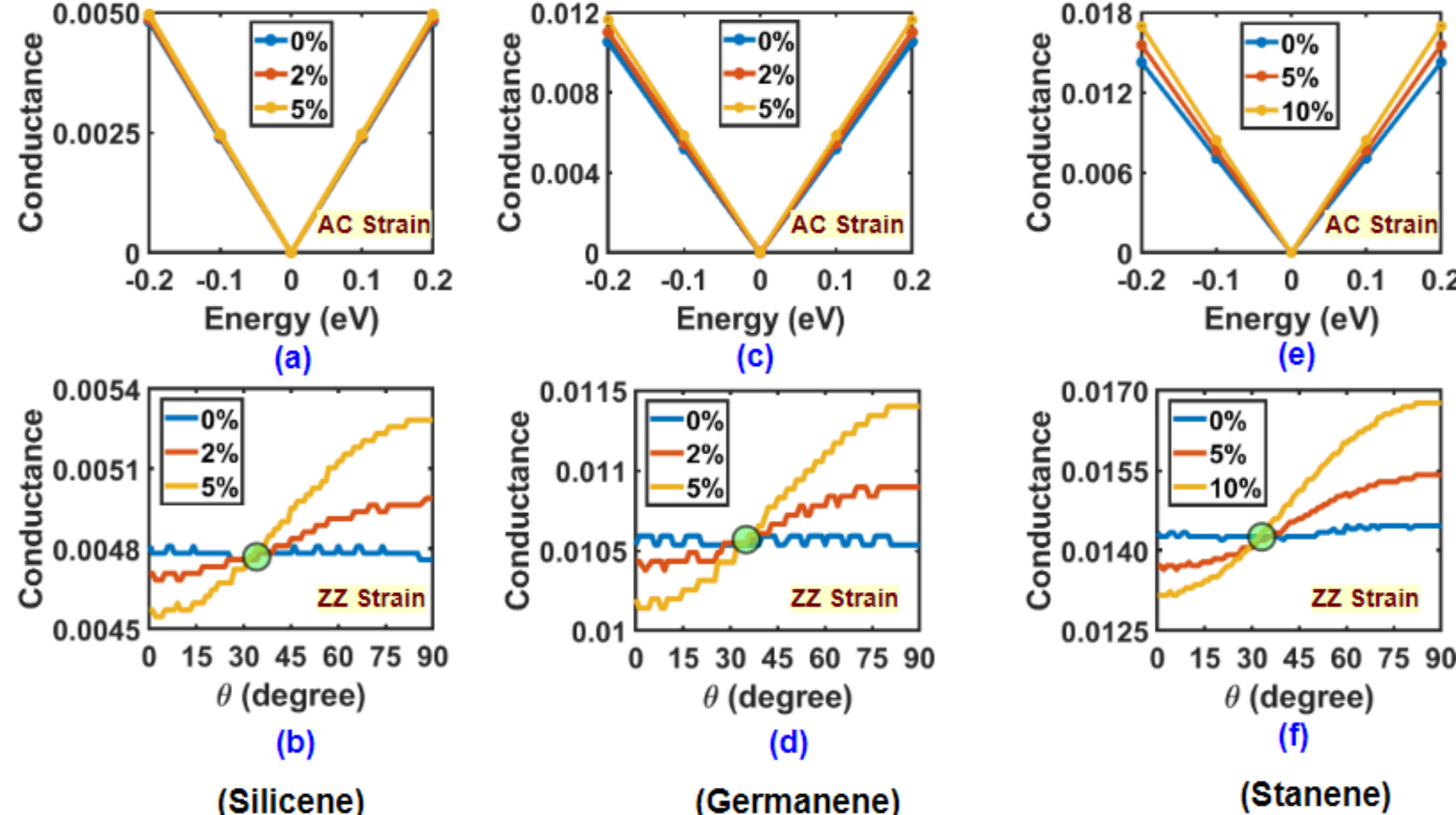}
     \caption{Plots of conductance as a function of energy (eV) for the applied strain having the values $0\%$,  $2\%$ and  $5\%$ for silicene, germanene and values  $0\%$,  $5\%$ and  $10\%$ for stanene along the (a),(c), (e) armchair direction. Plots of conductance as a  function of transport angle ($\theta$) and applied strain for the values $0\%$,  $2\%$ and  $5\%$ for silicene, germanene, and $0\%$,  $5\%$ and  $10\%$ for stanene along the (b), (d), (f) zigzag direction. The green circle in bottom row plots represent the strain insensitive transport angle.}
    \label{P02_8}
 \end{figure*}
Conductance modulation is the process of varying conductance concerning the strained transmission as a function of energy. This section aims to show the variation of conductance with respect to angular strain for buckled $2D$ Xenes. The transport properties of graphene taken from Landauer-Buttiker (LB) formalism~\cite{datta1997electronic} can serve as a template for its equivalent $2D$ Xenes, which are quasi-ballistic in nature. The corresponding ballistic conductance is given as $G(E)= G_0*T^{s}(E)$, where $G_0$ is quantum conductance, a fundamental parameter expressed as: $G_0=\frac{2e^2}{h}$. $T^{s}(E)$ represents strained transmission, defined in the previous section. \\
\indent As mentioned earlier, conductance is related to energy through transmission, so the plots for conductance concerning the energy window are shown in Fig.~\ref{P02_8}a, ~\ref{P02_8}c and ~\ref{P02_8}e with strain applied in AC direction. Similarly, plots for conductance as a function of $\theta$ are shown in Fig.~\ref{P02_8}b, ~\ref{P02_8}d and ~\ref{P02_8}f with strain applied in ZZ direction. These plots are applicable for monolayer buckled Xenes; however, puckered structures are also expected to have conductance modulation. Figures~\ref{P02_8}a,~\ref{P02_8}c and ~\ref{P02_8}e depict that conductance increases with an increase in energy and strain due to an increase in the number of transverse modes passing through the Dirac cone in AC strain. The increase in transverse modes is related to the deformation of the Dirac cone from circle to ellipsoid~\cite{sahoo2022silicene}. In the case of ZZ strain, the deformation will result in an oval shape which reduces the number of modes passing through the Dirac cone~\cite{sahoo2022silicene}. Hence, the plots will exhibit opposite characteristics in the ZZ strain. The increase in conductance for energy and strain is also observed in graphene~\cite{yan2021conductance}, which will validate our simulated plots. Variation of conductance with $\theta$ is shown in  Fig.~\ref{P02_8}b,~\ref{P02_8}d and ~\ref{P02_8}f for silicene, germanene and, stanene respectively. These plots resemble the plots of transmission and current. Innately, the conductance below and above the critical angle exhibits opposite behavior. Initially, conductance decreases with increased strain, and as we go beyond the critical angle (shown by the green circle), it starts increasing with strain. Zero strain conductance shows an almost flat line, but the small changes in the values are discretized. As the changes in plots are discrete, it is called quantized conductance.

\subsection{Connotation of the results}
Table~\ref{P02_table3} compares the piezoresistance GF of silicene, germanene, and stanene with other prominent materials like phosphorene, graphene, and popular semiconductors like silicon and germanium. From this table, we infer that the GF of silicene and germanene are petite, whereas their 3D counterparts, silicon, and germanium, have very high GF. Also, as we go along the periodic table, the GF of the monolayer buckled Xenes will increase. This increase in the value forms a pattern, which can be generalized to all the group $IV$ buckled Xenes. Phosphorene, a group $V$ element with puckered structure, has a very high GF, which can act as a frontispiece for other group $V$ elements.

\setlength{\tabcolsep}{16.5pt}
\begin{table}[h]
\caption{Comparisons of the GFs of various materials.}
\centering
\begin{tabular}{c c c}\\ 
 \hline\hline
 Material & GF & Reference\\[0.08cm]
 \hline
 Silicene & 0.758 & ~\cite{sahoo2022silicene}\\[0.03cm]
 Germanene & 1.391 & This work\\[0.03cm]
 Stanene & 2.349 & This work\\[0.03cm]
 Phosphorene & 120 & ~\cite{nourbakhsh2018phosphorene}\\[0.03cm]
 Graphene & 0.6 & \cite{sinha2020graphene}\\[0.03cm]
 Silicon & 200 & \cite{smith1954piezoresistance} \\[0.03cm]
 Germanium  & 150 &  \cite{smith1954piezoresistance} \\[0.08cm]
 \hline\hline
\end{tabular}
\label{P02_table3}
\end{table}

\begin{figure*}[t]
     \centering
     \includegraphics[width=0.95\textwidth]{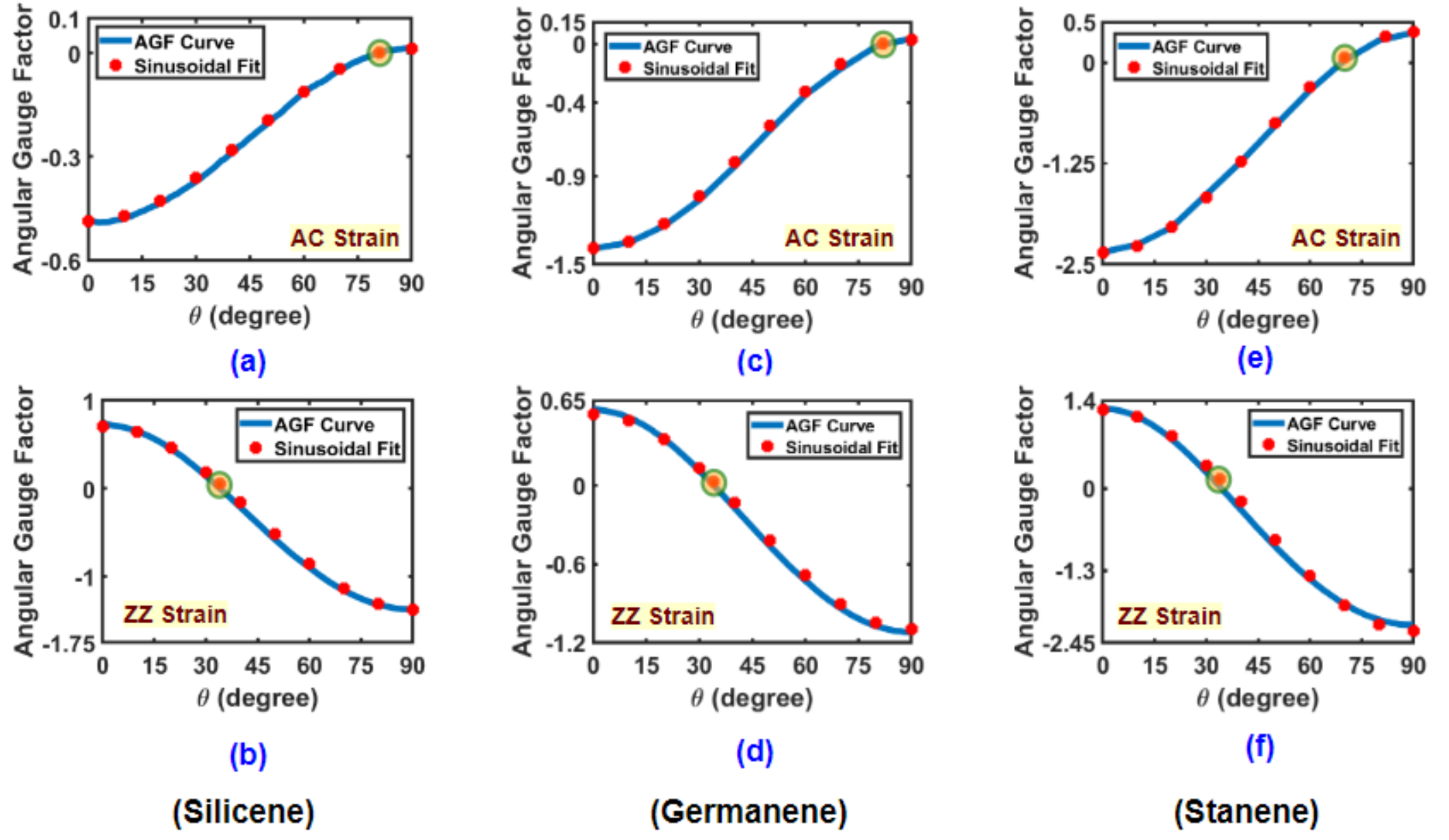}
     \caption{The plots of angular gauge factor along with its sinusoidal fit as a function of transport angle along the  (a),(c),(e) armchair and (b),(d),(f) zigzag directions for silicene, germanene, and stanene respectively. The highlighted point in each plot represents a critical angle.}
    \label{P02_9}
 \end{figure*}

 Figures~\ref{P02_9}a,~\ref{P02_9}c and ~\ref{P02_9}e and ~\ref{P02_9}b,~\ref{P02_9}d and ~\ref{P02_9}f depict AGFs as a function of the transport angle for silicene, germanene and stanene for AC and ZZ directions respectively. The AGFs have sinusoidal dependence on the angle ($\theta$) for both the AC and ZZ strains, and it is evident by now that the variation of AGF resembles a sinusoidal function similar to the one obtained by Sinha~\textit{et al.} for graphene because of the similarity in deformation of Dirac points in the reciprocal space. This variation in gauge factor goes from positive to negative in ZZ strain (see Fig.~\ref{P02_9}b,d,f) as the strained resistance is more than unstrained resistance(see Fig.~\ref{P02_8}b,d,f). The highlighted circle in each plot represents the critical angle denoting zero AGF, which stands for the angle insensitive to strain, and all the strained and unstrained parameters will have only one single value approximately. After the critical angle, the gauge factor becomes negative for ZZ strain.
 \setlength{\tabcolsep}{10.5pt}
\begin{table}[h]
  \caption{Values of constants P and Q for different Xenes to fit sinusoidal variation.}
 \centering
\begin{tabular}{c cc cc}
 \hline\hline
        &
 \multicolumn{2}{c}{\textit{AC Strain}} & \multicolumn{2}{c}{\textit{ZZ Strain}}\\
   Material  & P & Q & P & Q\\[0.1cm]
\hline
 Silicene & 0.249 & 0.237 & 1.079 & -0.239\\[0.05cm]
 Germanene & 0.710 & 0.268 & 0.820 & -0.275\\[0.05cm]
 Stanene & 1.376 & 1.009  & 1.7555  & -0.5075\\[0.1cm]
 \hline\hline
  \end{tabular}
\label{P02_table4}
  \end{table}
  
 In AC strain, the critical angle is the same ($81$$^{\circ}$) for both silicene and germanene, pointing out that, despite the difference in quantifying cell dimension and related aspects, still the two elements are very close to each other. For stanene, the critical angle is $69$$^{\circ}$, attributed to a change in applied strain and notable change in the parameters listed in table~\ref{P02_table5}. Interestingly, for ZZ strain, all the $3$ materials have the same critical angle of $34$$^{\circ}$ (ref Fig.~\ref{P02_9}(b, d and f)) despite all the differences, which draws an exciting conclusion for the rest of the group $IV$ elements. To fit the sinusoidal plot, mathematically, AGF is in the form of
 \begin{equation}
\mathrm{AGF= {-\{P \cos(2\theta)+Q\}}}, 
\end{equation}

 \setlength{\tabcolsep}{5.0pt}
  \begin{table*}
\caption{Comparison of various fundamental parameters of silicene, germanene, stanene, and phosphorene.}
\begin{ruledtabular}
\begin{tabular}{cccccccc}
 &\multicolumn{2}{c}{Silicene}&\multicolumn{2}{c}{Germanene}&\multicolumn{2}{c}{Stanene}&{Phosphorene}\\
 Parameters&$T(E)=1$&T(E) $<$ 1&$T(E)=1$&T(E) $<$ 1&$T(E)=1$&T(E) $<$ 1&$T(E)=1$
\\ [0.08cm]
\hline
 Dirac cone radius (1~meV) (\AA$^{-1}$)  & $0.0002928$ & $0.0002928$ & $0.0002872$ & $0.0002872$ & $0.0003252$ & $0.0003252$ & $0.0002159$ \\[0.03cm]
 Current ($\mu$A) &  $24.76$    & $4.12$ & $24.30$ & $9.11$ & $27.598$ & $12.26$  &  $5.5$~\cite{rout2019fundamentals}\\ [0.03cm]
 Resistance (K$\Omega$) & $0.404$   & $2.428$  & $0.409$ & $1.092$ & $ 0.360$ & $0.815$ & -\\
 Conductance($\mho$) &  $0.0288$ & $0.00481$ & $0.03148$ & $0.0111$ & $0.03226$ &$0.01673$ & $0.00061$~\cite{rout2019fundamentals}\\
\end{tabular}
\end{ruledtabular}
\label{P02_table5}
\end{table*}

where P and Q are constants, and their values for buckled Xenes are listed below in the table~\ref{P02_table4}. Several combinations can be possible for the constants, but the listed values suit the best for sinusoidal fit regarding the variation of AGF concerning the transport angle. \\

\indent Table~\ref{P02_table5} lists various fundamental parameters of buckled and puckered Xenes like silicene, germanene, stanene, and phosphorene with and without transmission probability at zero strain. Transport in a ballistic regime is denoted as $T(E)=1$, stating it is independent of the mean free path, and transmission for a quasi-ballistic regime is denoted as $T(E)<1$. The former is called without transmission probability, and the latter is called with transmission probability.
Dirac cone radius is invariant to transmission probability as it is deprived of any shifting and deformation at zero strain. So, the radius will hold its value irrespective of the above-mentioned regimes. However, the radius of the Dirac cones increases as the atomic number of the constituent group $IV$ element increases, starting from graphene to stanene at zero strain. Considering transmission to be equal to one, the current value will increase. The corresponding effects can be seen for resistance and conductance. We obtain the value of resistivity as \SI{0.404}{\kilo\ohm}, \SI{0.409}{\kilo\ohm}, \SI{0.360}{\kilo\ohm} for buckled Xenes in ballistic regime. The resistivity value for stanene, germanene, and silicene is less compared to the resistivity of graphene~\cite{sinha2019piezoresistance}. This small resistivity value is because the number of TMs in other buckled Xenes is more, than in graphene (for the sheet width) due to the larger size of its Dirac cones. For phosphorene, we have considered only the ballistic regime. Its Dirac cone radius is the smallest of all the listed values, resulting in the minimum current value compared to other corresponding values taken only in the ballistic regime. We have observed anisotropy in the shifting and deformation of the Dirac cone for phosphorene in the presence of strain either in AC or ZZ direction. The quantitative results are also listed (see Supplementary information) for clarity. The proportion between the parameters, irrespective of transmission probability, is maintained throughout, which persuades the correctness of the result. \\
\indent  The discussed buckled Xenes are atomically-thin membranes as they are monolayers and are expected to have a high value of adhesivity (like most of the similar 2D materials)~\cite{megra2019adhesion} and elasticity~\cite{peng2013mechanical}. These significant values lead to a future endeavor, where these Xenes are expected to have a very high-pressure sensitivity like planar monolayer graphene ~\cite{sinha2022ballistic} and $\mathrm{PtSe_{2}}$~\cite{smith2013electromechanical, wagner2018highly} despite a low GF. These Xene sheets also show high resistance to the change in strain. This resistant is realized when the normalized resistance is plotted against strain for silicene and graphene ~\cite{sahoo2022silicene}, which unveils their vital electronic characteristic to the change in strain value. Several characteristics like levelheaded conductivity~\cite{zhang2014thermal}, high mobility calculation~\cite{mir2020recent}, and high elastic limit~\cite{peng2013mechanical} of buckled Xenes will enable them to act as electrodes and interconnects in flexible electronic devices. Also, they have the persuasion to act as potential topological materials~\cite{zhao2020two}. Based on results obtained in this voluminous model, we propound that the Xenes can be strong contenders in flexible electronics and show promising characteristics in any application related to straintronics.

\section{Conclusion} \label{section_4}
In this paper, we investigated the piezoresistance of buckled Xenes like germanene and stanene using \textit{ab-initio} density function theory and quantum transport theory. We compared the result with silicene and graphene. We calculated the directional piezoresistance for different strain values along the armchair and zigzag directions. We typically obtained a smaller value of the directional piezoresistances and their sinusoidal dependence. The value of the gauge factor increases as we go up for group $IV$ elements along their atomic numbers in the periodic table. The strain-insensitive transport angles corresponding to the zero gauge factors are $81^{\circ}$ and $34^{\circ}$ for armchair and zigzag strains, respectively, for silicene and germanene. For stanene as the strain limit is extended to $10\%$ and notable changes in the fundamental parameters, the critical angle for stanene along armchair and zigzag directions are $69^{\circ}$ and $34^{\circ}$ respectively. The small gauge factor of buckled Xenes can be attributed to its robust Dirac cone and strain-independent valley degeneracy. We have realized conductance quantization in a quasi-ballistic regime for the buckled Xenes. We have also shown the strained tight binding parameters of phosphorene, which is the first of its kind in the case of puckered Xenes. Further, we are exploring the spintronics part of $2D$ Xenes, where we will see the response of the interfacial effects between the contacts and the Xenes against the applied strain. Based on the obtained results, we propose the buckled Xenes as an interconnect in flexible electronics and are promising candidates for various applications in straintronics.\\

\begin{acknowledgements}
We thank Dr.Abhinaba Sinha for his guidance and support towards preparing this manuscript. The Research and Development work undertaken in the project under the Visvesvaraya Ph.D. Scheme of Ministry of Electronics and Information Technology, Government of India, is implemented by Digital India Corporation (formerly Media Lab Asia). This work was also supported by the Science and Engineering Research Board (SERB), Government of India, Grant No. CRG/2021/003102.
\end{acknowledgements}
\bibliography{reference}
\end{document}


\maketitle

The shifting of K-point away from $\Gamma$-point in AC strain and towards $\Gamma$-point in ZZ strain for silicene, germanene and stanene is quantified in the following table. The shifting of K-point is denoted by another point called Dirac point (DP). In AC strain, DP magnitude is higher when K-point shift towards M-point and opposite happens for ZZ strain which will reflect in the relative difference between K-point and DP-point. 

\setlength{\tabcolsep}{6.3pt}
\begin{table}[h]
    \centering
\begin{tabular}{c cc cc cc}
 \hline\hline
        &
    \multicolumn{2}{c}{Silicene} & \multicolumn{2}{c}{Germanene} & \multicolumn{2}{c}{Stanene} \\
        Strain & AC strain & ZZ strain & AC strain & ZZ strain & AC strain & ZZ strain \\[0.1cm]
        
        values & K-DP & K-DP & K-DP & K-DP & K-DP & K-DP \\[0.1cm]
\hline
 $0\%$ & $0$ & $0$ & $0$ & $0$ & $0$ & $0$ \\[0.05cm]
 $2\%$ &  $-0.0051$ & $0.0064$  & $-0.0032$  & $0.0036$ & $-0.0017$ & $0.0027$  \\[0.05cm]
 $5\%$ &  $-0.0129$ & $0.0170$  & $-0.0082$  & $0.0089$ & $-0.0042$ & $0.0074$  \\[0.1cm]
 \hline\hline
  \end{tabular}
 \caption{Shifting of Dirac cones from the K-points (in \AA$^{-1}$) for armchair and zigzag strain in case of silicene, germanene and stanene respectively.}
\label{P02_table_S1}
  \end{table}

The pictorial representation of calculation of mode density is shown below. Stanene, is taken for example where the $10\%$ strain is applied along both AC and ZZ direction denoted by $S_x$ and $S_y$ respectively. We have also shown the deformation of Dirac cone with applied strain which is in good agreement with the drawn figure. 

\begin{figure}[h]
\centering
\includegraphics[scale=0.55]{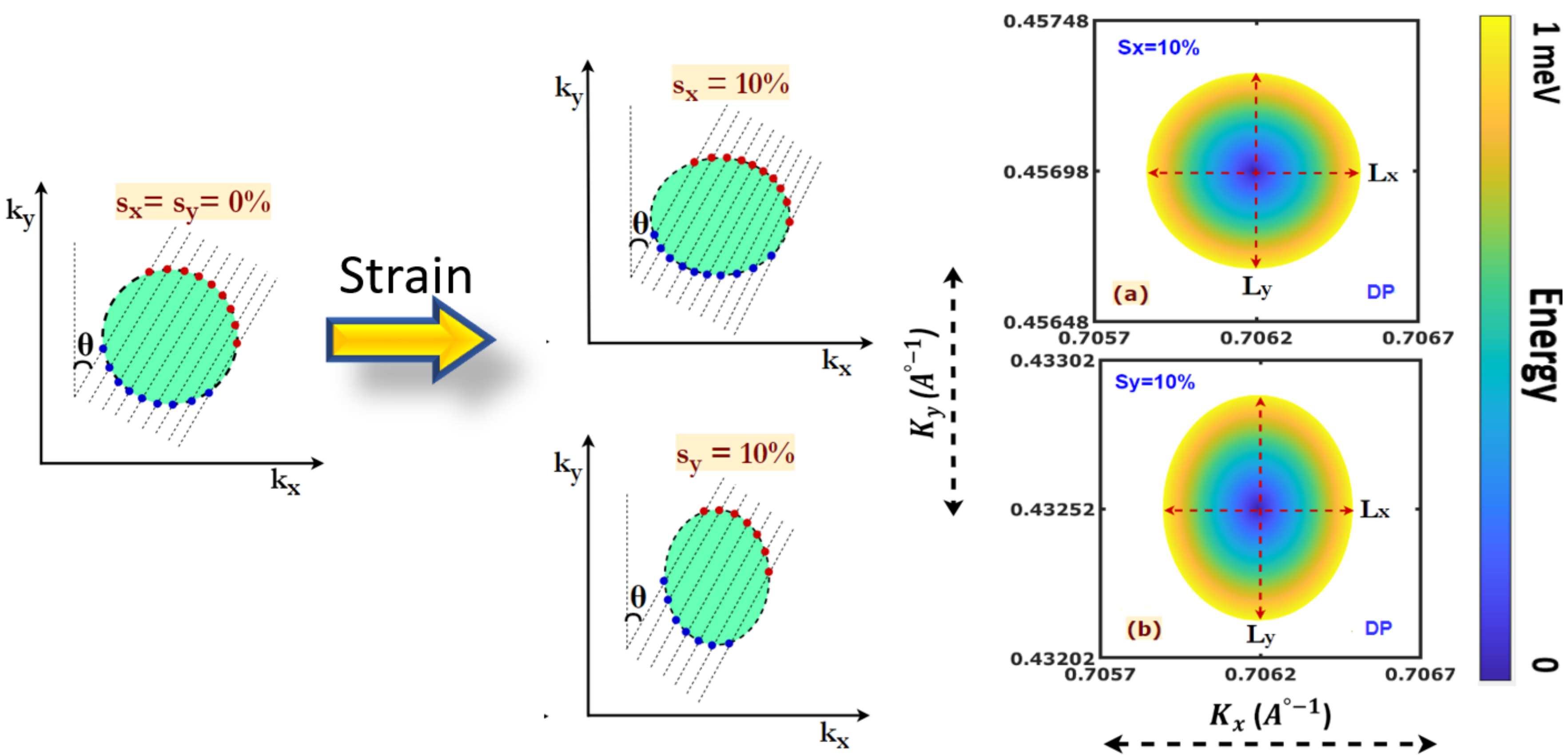}
\caption{Schematic diagram depicting the modes in constant energy surfaces at 0\% strain, 10\% strain along the AC and ZZ direction. The dotted lines represent the direction of transverse modes, and the red and blue dots represent the modes for forward and backward propagating electrons respectively. Energy color-map near the Dirac point of stanene with $S_x$=$S_y$=$10\%$.}
\label{fig:P02_S1}
\end{figure}

In order to support the above diagram, the deformation of stanene, can also be quantified and it is listed in the following table. 

\setlength{\tabcolsep}{12.3pt}
\begin{table}[h]
    \centering
\begin{tabular}{c cc cc}
 \hline\hline
        &
 \multicolumn{2}{c}{\textit{AC Strain    }} & \multicolumn{2}{c}{\textit{ZZ Strain}}\\
        Strain ($\%$)  & $DP (L_x)$ & $DP (L_y)$ & $DP (L_x)$ & $DP (L_y)$\\[0.1cm]
\hline
 $0$ & $0.0006505$ & $0.0006505$ & $0.0006505$ & $0.0006505$\\[0.05cm]
 $5$ & $0.0007009$ & $0.0006401$ & $0.0006204$ & $0.0007002$\\[0.05cm]
 $10$ & $0.0007584$ & $0.0006307$  & $0.0005931$  & $0.0007561$\\[0.1cm]
 \hline\hline
  \end{tabular}
 \caption{The major and minor axes of the deformed Dirac cones (in \AA$^{-1}$) at E$=1$ $meV$  for armchair and zigzag strain in case of stanene.}
\label{P02_table_S2}
  \end{table}

For, phosphorene considering the shifting of Dirac points, table below gives the quantitative picture. Taking $X$-point as $DP_1$ , $S$-point as $DP_2$, $Y$-point as $DP_3$ and $\Gamma$-point as $DP_4$ we will list down one dimensional shifting of $DP1$ and $DP3$  with respect to strain applied along AC and ZZ strain. For $DP_1$, shifting with respect to x-axis and for $DP_3$, shifting along y-axis is shown which shows clear anisotropy in phosphorene. Again the deformation can also be shown in table.~\ref{P02_table_S4} which show clear anisotropy.

\setlength{\tabcolsep}{12.3pt}
\begin{table}[h]
    \centering
\begin{tabular}{c cc cc}
 \hline\hline
        &
 \multicolumn{2}{c}{\textit{AC Strain    }} & \multicolumn{2}{c}{\textit{ZZ Strain}}\\
        Strain ($\%$)  & $DP_1$ & $DP_3$ & $DP_1$ & $DP_3$\\[0.1cm]
\hline
 $0$ & $0$ & $0$ & $0$ & $0$\\[0.05cm]
 $5$ & $-0.0076$ & $0.1328$ & $0.031$ & $-0.0956$\\[0.05cm]
 $10$ & $-0.0155$ & $0.2601$  & $0.0625$  & $-0.193$\\[0.1cm]
 \hline\hline
  \end{tabular}
 \caption{Shifting of Dirac cones from their respective K-points (in \AA$^{-1}$) for armchair and zigzag strain in case of phosphorene.}
\label{P02_table_S3}
  \end{table}

\setlength{\tabcolsep}{12.3pt}
   \begin{table}[h]
    \centering
\begin{tabular}{c cc cc}
 \hline\hline
        &
 \multicolumn{2}{c}{\textit{AC Strain    }} & \multicolumn{2}{c}{\textit{ZZ Strain}}\\
        Strain ($\%$)  & $DP_1 (L_x)$ & $DP_1 (L_y)$ & $DP_3 (L_x)$ & $DP_3 (L_y)$\\[0.1cm]
\hline
 $0$ & $0.0001379$ & $0.0001379$ & $0.0004264$ & $0.0004264$\\[0.05cm]
 $5$ & $0.0001489$ & $0.0001286$ & $0.0002980$ & $0.0004480$\\[0.05cm]
 $10$ & $0.0001605$ & $0.0002144$  & $0.0002411$  & $0.0004716$\\[0.1cm]
 \hline\hline
  \end{tabular}
 \caption{The major and minor axes of the deformed Dirac cones like $DP_1$ AND $DP_3$ (in \AA$^{-1}$) at E$=1$ $meV$  for armchair and zigzag strain for phosphorene.}
\label{P02_table_S4}
  \end{table}